\documentclass[prd,article,nofootinbib,twocolumn,preprintnumbers,superscriptaddress]{revtex4}
\usepackage{graphicx,amsfonts,color,comment,amsmath,hyperref,float,times}
\usepackage{amssymb}	
\usepackage{graphicx,amsfonts,color,comment,amsmath,hyperref,float}
\usepackage{amssymb}	
\usepackage{ragged2e}

\usepackage{mathrsfs,amssymb}  

\newcommand{\be}{\begin{equation}}
\newcommand{\ee}{\end{equation}}
\newcommand{\bea}{\begin{eqnarray}}
\newcommand{\eea}{\end{eqnarray}}

\begin{document}

\title{A COHERENT Search Strategy for Beyond Standard Model Neutrino Interactions}

\author{Ian M. Shoemaker}
\affiliation{Department of Physics, University of South Dakota, Vermillion, SD 57069, USA}

\date{\today}
\begin{abstract}
We study the sensitivity of the stopped pion source of neutrinos at the Spallation Neutron Source (SNS) at Oak Ridge National Laboratory to neutrino non-standard interactions (NSI). In particular, we find that simple event counting above threshold can improve constraints on NSI for electron- and muon- flavored NSI, with the strongest constraints arising for flavor-diagonal NSI coupling to $\mu$ neutrinos.  However, if the detector resolution is sufficient to all for even a coarse spectral study of events, COHERENT will also be sensitive to the mass scale of NSI. We demonstrate that this can yield new limits on $Z'$ completions of NSI if the gauge boson mass is $\lesssim 1$ GeV.  

\end{abstract}
\preprint{}


\maketitle

\section{Introduction}

Neutrinos remain poorly constrained. With the possible exception of the Higgs, they are arguably the least understood particles of the Standard Model (SM). Simply as an empirical matter then, they may easily contain signs of new physics beyond the SM (BSM). Indeed, with the origin of neutrino masses remaining unknown, it is clear that BSM physics affects neutrinos at some level.  Yet despite this theoretical possibility and rapid experimental progress, even today neutrino interactions remain one of the least tested predictions of the SM.  Clearly such a test will be extremely fruitful whether it leads to additional constraints and a confirmation of the SM, or more optimistically, to a discovery leading the path the new physics. 

In this paper we will focus on the neutral current (NC) class of such interactions that can imprint themselves on both neutrino propagation and scattering, dubbed ``non-standard neutrino interactions'' (or NSI). NSI has a long history and has been studied for as long as the matter effect on neutrino oscillations has been known~\cite{Wolfenstein:1977ue}. 

The strength of these new neutrino interactions is parametrized in units of the Fermi constant $G_{F}$, 
\be 
\mathscr{L}_{{\rm NSI}} \supset -2\sqrt{2} G_{F} \sum_{f,P,\alpha,\beta} \varepsilon_{\alpha \beta}^{f,P}\left(\bar{\nu}_{\alpha}\gamma^{\mu}\nu_{\beta}\right)\left(\bar{f}\gamma_{\mu}Pf\right),
\label{eq1}
\ee
where $P$ is either the left/right chirality projection operator, $f$ is a SM fermion of ordinary matter ($e,u,d$), and the $\varepsilon$'s dictate the strength of the interaction. Notice that in contrast with the SM, NSI allows for the possibility of neutrino flavor off-diagonal NC couplings.

Although some of the $\varepsilon$ coefficients can be very strongly constrained, recent work has emphasized that light, weakly coupled vector mediators can produce surprisingly strong NSI~\cite{Friedland:2011za,Farzan:2015doa,Franzosi:2015wha,Farzan:2015hkd,Farzan:2016wym}. Note that although $Z'$ models are a natural way of UV completing the operators in Eq.~(\ref{eq1}), other models exist as well. For example upon Fierz rearrangement, one can show that scalars coupling as $\bar{f} \phi L$ will also produce NSI operators~\cite{Bergmann:1999pk,Berezhiani:2001rs,Wise:2014oea,Forero:2016ghr}. Models of this type include leptoquarks, and $R$-parity violating supersymmetry. However present constraints from electroweak precision data and rare decays are constrain such NSIs to be $\lesssim 1\%~G_{F}$ in strength~\cite{Friedland:2011za}. Therefore we will focus on $Z'$ models as NSI completions (though see  Ref.~\cite{Forero:2016ghr} for recent NSI model-building using an electrophilic Higgs).

Intriguingly, present current solar neutrino data may favor a non-standard matter potential which either NSI~\cite{Miranda:2004nb,Palazzo:2011vg,Friedland:2012tq} or dark matter-neutrino interactions may explain~\cite{Capozzi:2017auw}. This makes orthogonal probes an appealing test. In the dark matter case, additional probes are limited to future IceCube and better observations of small-scale DM structure. In the NSI case however, there are a variety of complementary search strategies. These include missing energy and multi-lepton searches at colliders~\cite{Berezhiani:2001rs,Davidson:2011kr,Friedland:2011za,Franzosi:2015wha}, long-baseline oscillation data~\cite{Wolfenstein:1977ue,Friedland:2012tq,Coelho:2012bp,Masud:2015xva,Sousa:2015bxa,deGouvea:2015ndi,Coloma:2015kiu,Masud:2016bvp}, cosmological considerations~\cite{Mangano:2006ar,Gonzalez-Garcia:2016gpq}, and neutrino-nucleus scattering data~(see e.g.~\cite{Davidson:2003ha,Scholberg:2005qs,Barranco:2005yy,Ohlsson:2013vaa,Dutta:2015vwa,Lindner:2016wff}). And finally, the sensitivity of atmospheric~~\cite{Fornengo:2001pm,Guzzo:2001mi,GonzalezGarcia:2004wg,Friedland:2004ah,Friedland:2005vy,GonzalezGarcia:2011my,Mocioiu:2014gua} and solar data~~\cite{Friedland:2004pp,Bolanos:2008km,Palazzo:2009rb,Palazzo:2011vg,Bonventre:2013loa,Gonzalez-Garcia:2013usa,Farzan:2015doa,Maltoni:2015kca} to NSI has been extensively studied as well. 

In the near-term, the COHERENT experiment utilizing the SNS at Oak Ridge National Laboratory (ORNL) will attempt to measure coherent elastic neutrino-nucleus (CE$\nu$NS) scattering for the first time~\cite{Akimov:2015nza}. Despite the large cross section this process should have, it has yet to be measured because of the low-energy thresholds needed in order to observe such nuclear recoils. The notion of constraining NSI with a stopped pion source was first emphasized in~\cite{Scholberg:2005qs}. Note that COHERENT will also have novel sensitivity to light sterile neutrinos~\cite{Kosmas:2017zbh}, $Z'$ bosons~\cite{Barranco:2007tz}, and models of light dark matter~\cite{deNiverville:2015mwa}. 

The Spallation Neutron Source uses stopped pions to produce a high-intensity flux of neutrinos. Thus the flavor composition is 2:1 in favor of muon-type neutrinos, with a prompt muon-flavored monochromatic component via
\be \pi^{+} \rightarrow \mu^{+} + \nu_{\mu}
\ee
and a more spread-out distribution of electron- and muon-flavored neutrinos via
\be
\mu^{+} \rightarrow \bar{\nu}_{\mu} + \nu_{e} + e^{+}
\ee
%


%
%

The outline of this paper is as follows.  First, we discuss our mock-up of the COHERENT detector, treatment of NSI, and set-up of our analysis. We then present the results of our analysis, focusing on the ability of COHERENT to provide both improved quantitative NSI sensitivity but also highlighiting the new qualitative sensitivity to the mass-scale of NSI. We then conclude.  

\section{COHERENT Detector Simulation Details}

\begin{figure*}[t]
\begin{center}
\includegraphics[width=.41\textwidth]{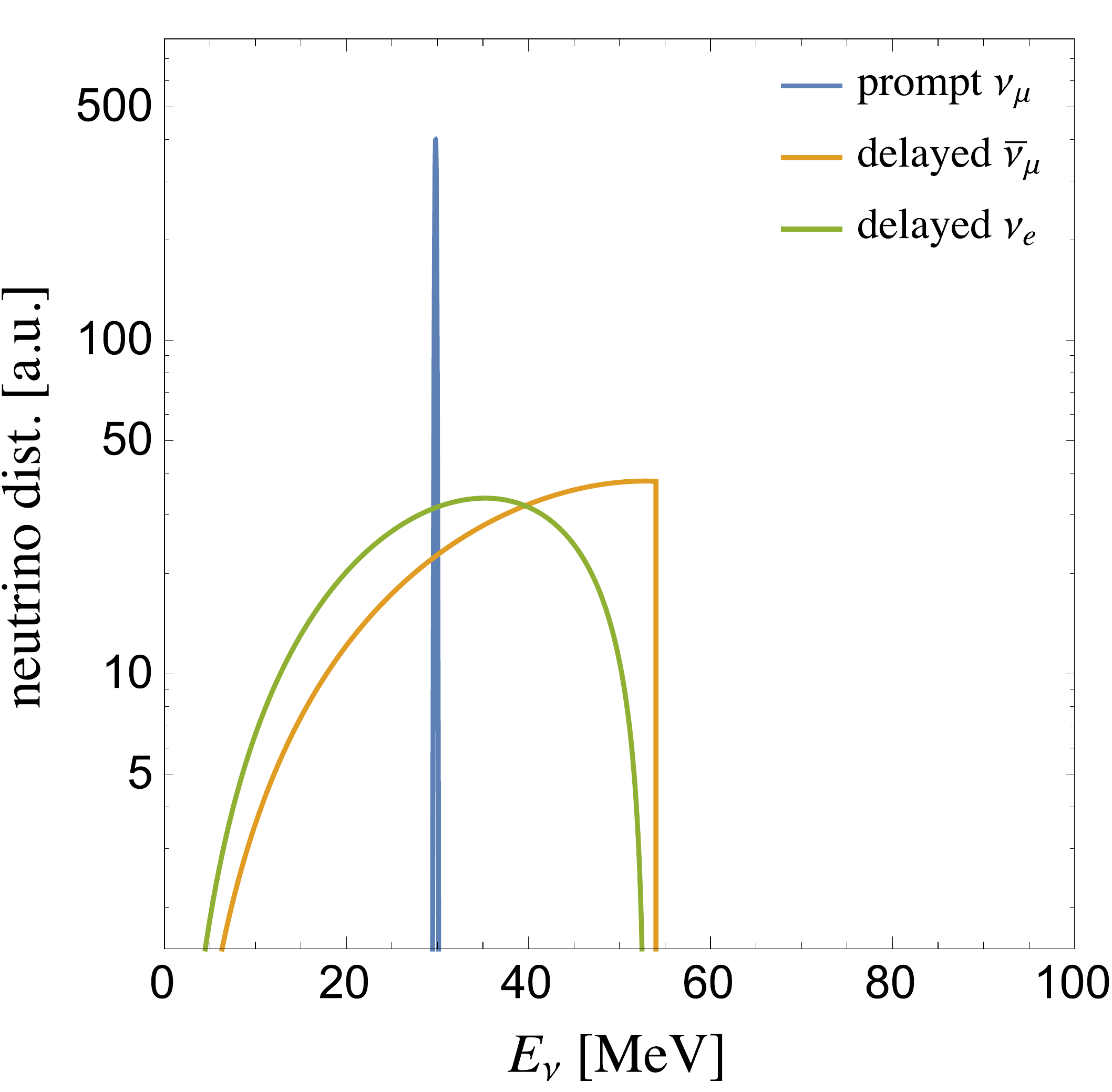}
\includegraphics[width=.4\textwidth]{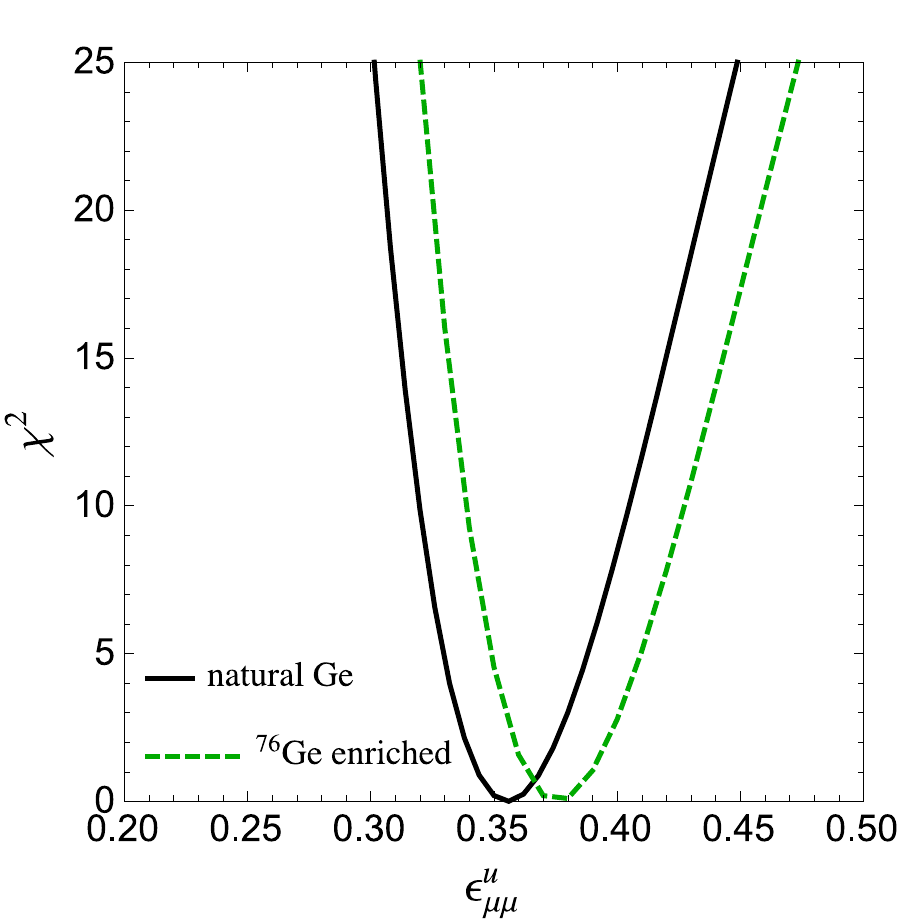}
\caption{{\it Left:}The dominant contributions to the prompt signal come from the two-body decay of the $\pi^{+}$ yielding a monochromatic $\nu_{\mu}$, while the delayed signal is dominated by the $\mu^{+}$ resulting in the spectra of $\nu_{e}$'s and $\bar{\nu}_{e}$'s. {\it Right:}The presence of a distribution of naturally occurring Ge isotopes aides somewhat in breaking the degeneracy discussed in the text. }
\label{fig1}
\end{center}
\end{figure*}
\subsection{Details of Mock-up}

In general NSI can have 6 new parameters to describe matter-neutrino interactions. However, let us for simplicity turn on only one new NSI coupling for illustrative purposes, the coupling to up-type quarks coupling only diagonally to neutrino flavor. Thus in other words all NSI couplings are zero except, $\varepsilon_{\alpha \alpha}^{uV}\neq 0$.

The generalized cross section for a $\nu_{\alpha}$ incident on a nucleus $\mathcal{N}$ is 
\be 
\frac{d \sigma_{\alpha\mathcal{N}}}{dE_{{\rm R}}}  = \frac{G_{F}^{2}m_{\mathcal{N}}}{\pi} F_{\mathcal{N}}^{2}(E_{{\rm R}}) \left(1- \frac{m_{\mathcal{N}}E_{{\rm R}}}{2E_{\nu}^{2}}\right) \times Q_{\alpha \mathcal{N}}^{2}
\ee
where the effective charge is

%
\be
 Q_{\alpha \mathcal{N}}^{2} \equiv \left[ Z  \tilde{g}_{p}^{V} + (A-Z) \tilde{g}_{n}^{V}  \right]^{2}
\ee
and the nucleon-level couplings are
\bea
\tilde{g}_{p}^{V} &=& g_{p}^{V} + 2 \varepsilon_{\alpha \alpha}^{uV}\\ 
\tilde{g}_{n}^{V} &=& g_{n}^{V} +  \varepsilon_{\alpha \alpha}^{uV}
\eea
where the SM couplings are $g_{n}^{V} = -1/2$, $g_{p}^{V} = 1/2 - 2 \sin^{2} \theta_{w}$. The analogous effective couplings for NSI $d$-type or universal quark couplings are straightforwardly obtained.

More generally however, we will be interested in the case that the mediator inducing NSI is comparable to or even lighter than the mediator's mass. In this case the effective charge is energy dependent  $Q_{\alpha \mathcal{N}}^{2} \rightarrow  Q_{\alpha \mathcal{N}}^{2} (E_{R})$, since the propagator of the $Z'$ enters. This leads to the potential of the being able to deduce the mediator mass if spectra are available. We allow for this more general possibility in our analysis by introducing an energy-dependent charge:
\bea
\tilde{g}_{p}^{V}&=&  g_{p}^{V} + \frac{2 g^{2}G_{F}^{-1}}{q^{2}+M_{Z'}^{2}} \\
\tilde{g}_{n}^{V}&=& g_{n}^{V} + \frac{g^{2}G_{F}^{-1}}{q^{2}+M_{Z'}^{2}}
\eea
where the momentum transfer is $ q^{2}  = 2 m_{\mathcal{N}} E_{{\rm R}}$.

\begin{figure*}[t]
\includegraphics[width=.3\textwidth]{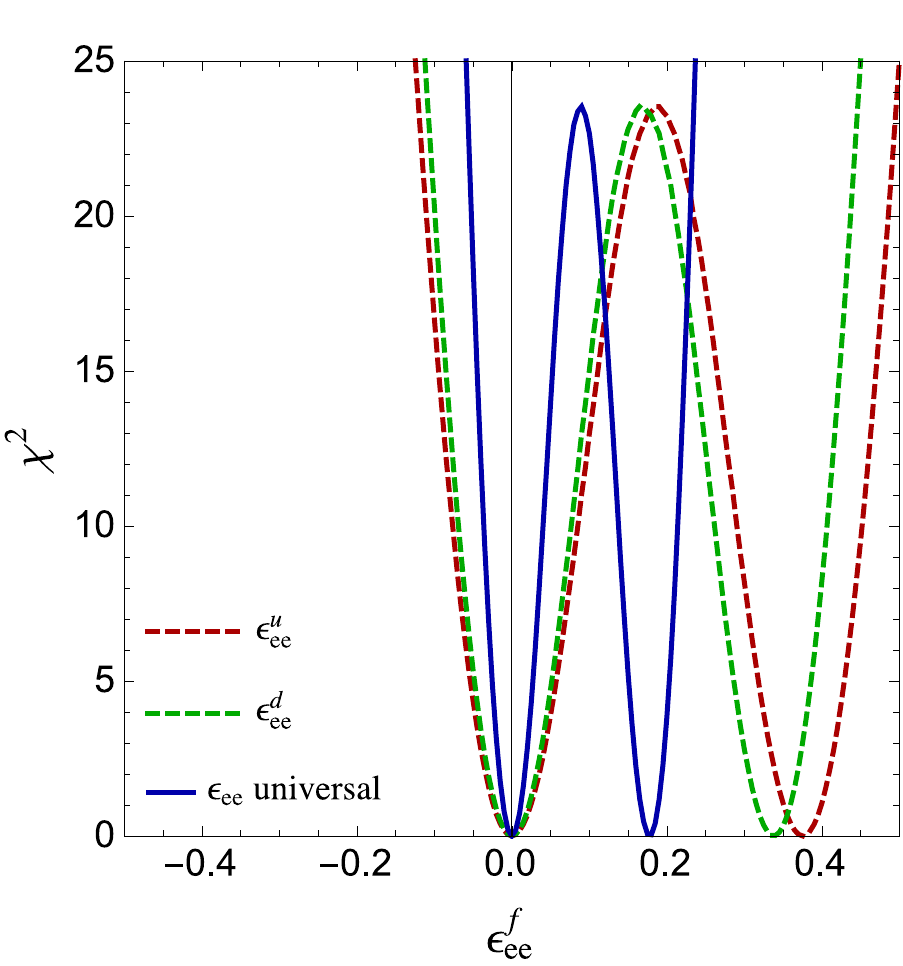}
\includegraphics[width=.3\textwidth]{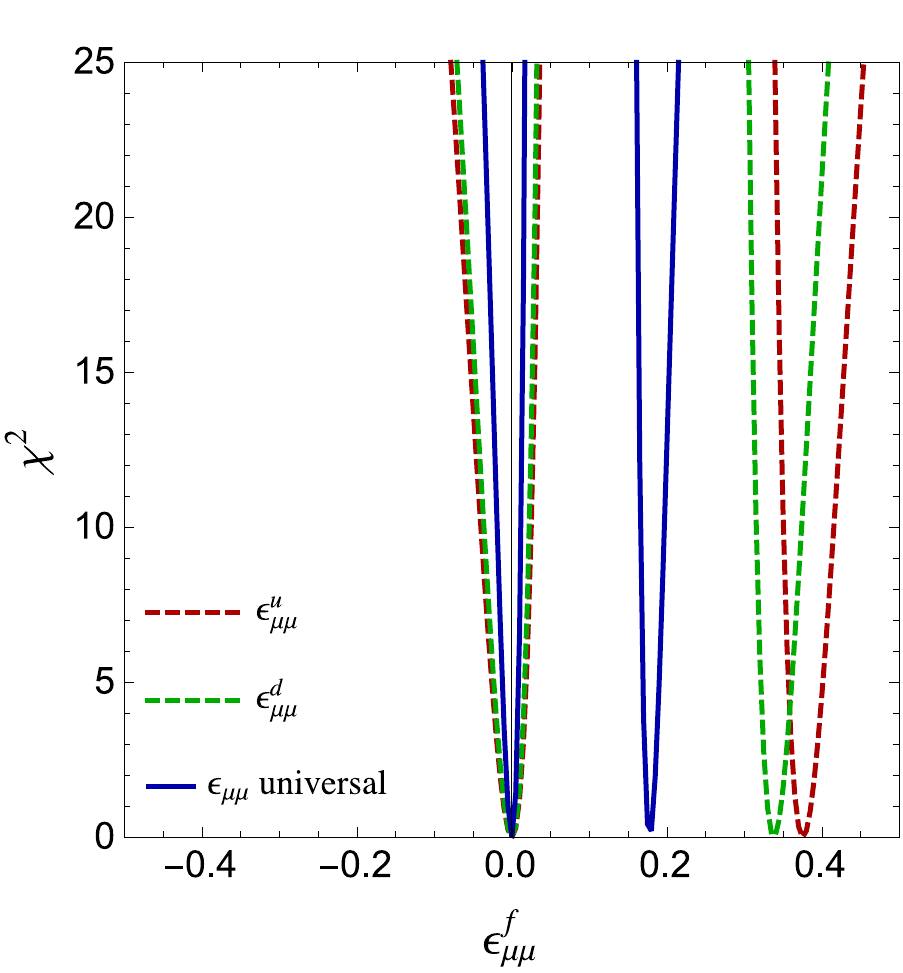} 
\includegraphics[width=.3\textwidth]{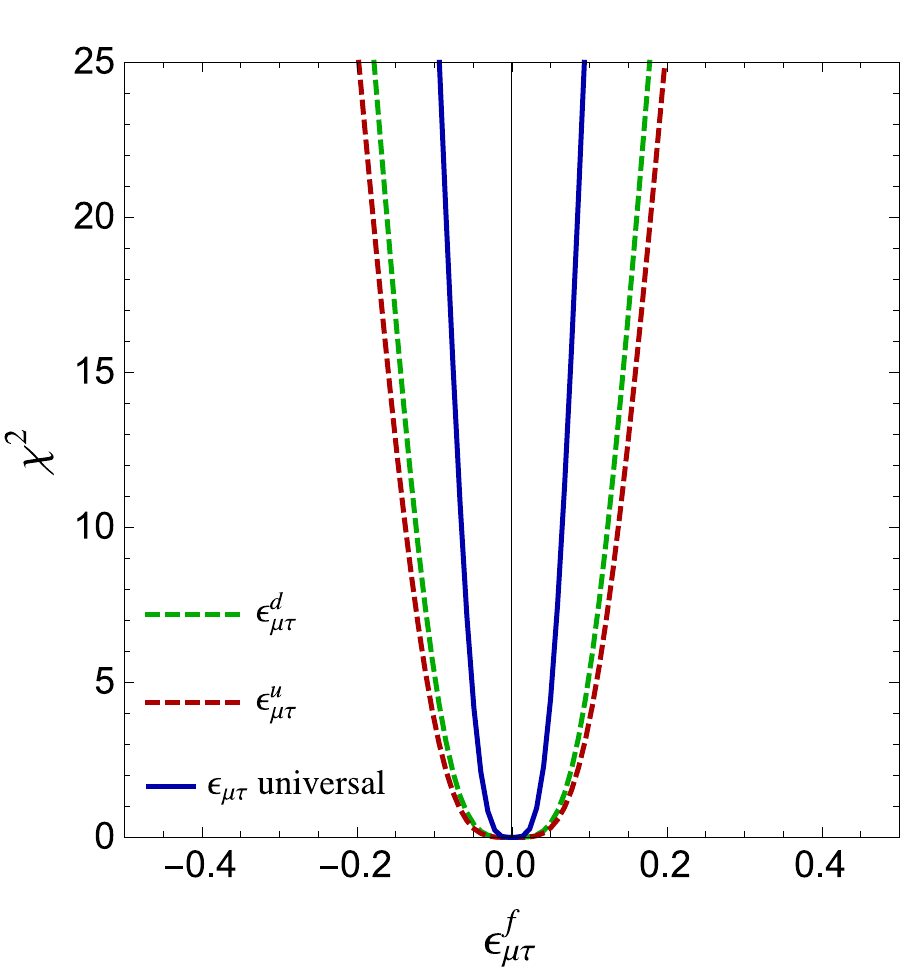}
\includegraphics[width=.3\textwidth]{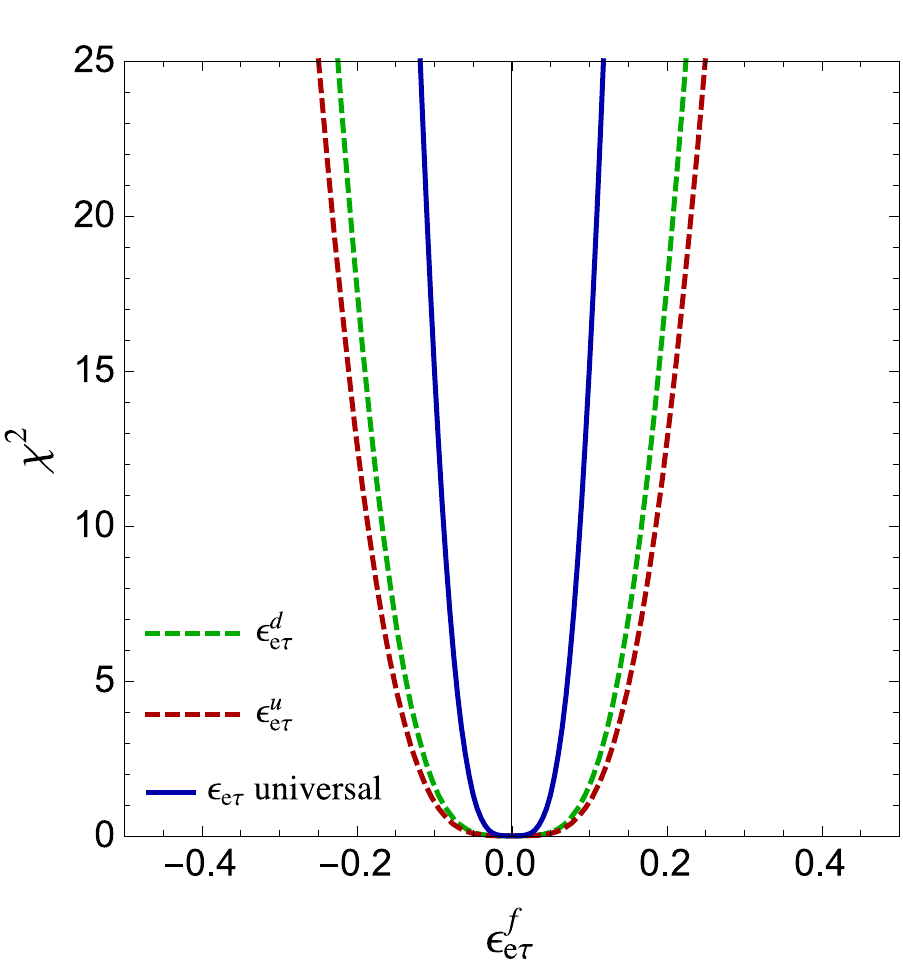}
\includegraphics[width=.3\textwidth]{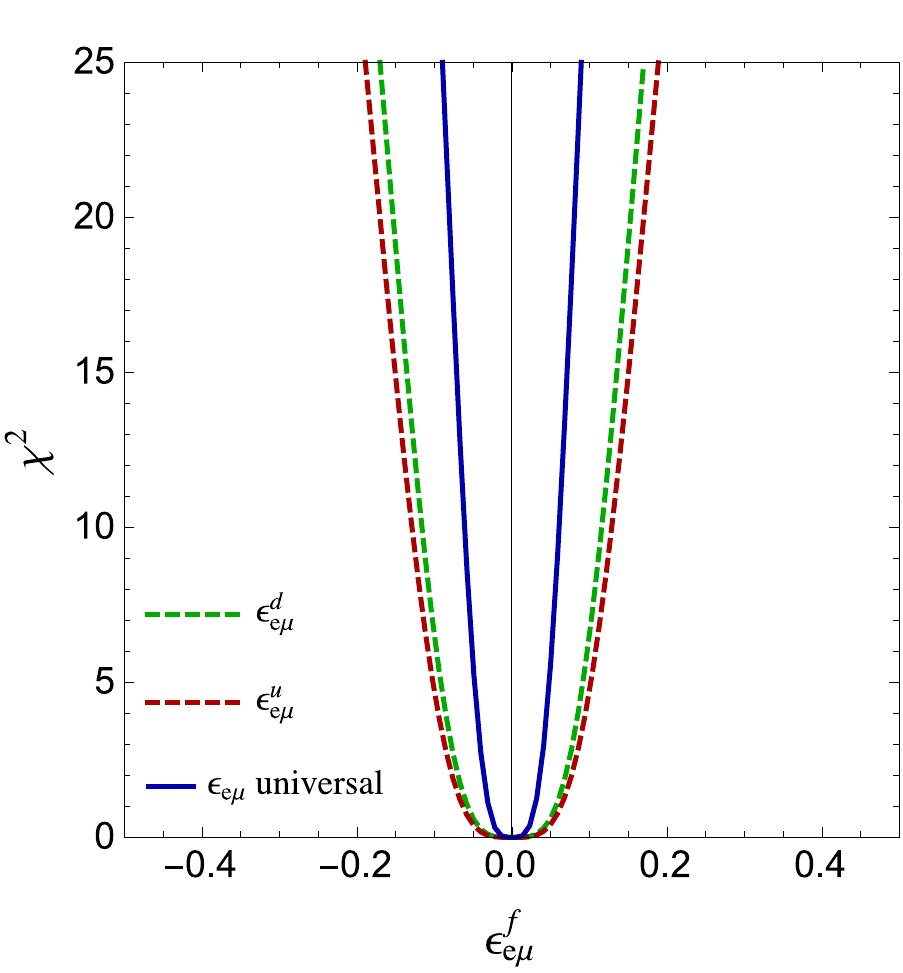}
\caption{Here we display the projected sensitivity of COHERENT to various NSI couplings. The red, green, and blue curves for each panel show the limit obtained from up-quark only coupling, down-quark only coupling, and universal quark couplings respectively.  Each panel displays the sensitivity for different NSI neutrino flavor couplings.  Note that the flavor off-diagonal terms are much less constrained since they do not benefit from interference with the SM. Relatedly the $\epsilon_{\mu\mu}$ NSI coupling is most strongly constrained since it benefits from both the interference effect and the larger $\nu_{\mu}$ flux. COHERENT does not have sensitivity to $\epsilon_{\tau\tau}$ NSI.}
\label{fig:pheno}
\end{figure*}


The number of events per unit detector mass per unit time induced by each neutrino flavor is
\be
\frac{dR_{\alpha}}{dE_{{\rm R}}} = \int \phi_{\alpha} \frac{d\sigma_{\alpha}}{dE_{{\rm R}}} dE_{\nu}
\ee
where the spectral shapes of each flux $\phi_{\alpha}$ are easily calculated from the kinematics of $\pi^{+} \rightarrow \mu^{+} + \nu_{\mu}$ and $ \mu^{+} \rightarrow \bar{\nu}_{\mu} + \nu_{e} + e^{+}$. For reference we display these in the left panel of Fig.~\ref{fig1}.

The prompt and delayed signals can be mostly separated from each other by a timing cut, though there is still a finite probability for the muon to decay within the proton pulse. Consistent with other work we take this probability to be $P_{\tau} = 0.14$~\cite{Coloma:2017egw}. 

Then we compute the predicted prompt and delayed events via this timing cut as
\bea 
N_{{\rm D}} &=& (1-P_{\tau}) (N_{\nu_{e}} + N_{\bar{\nu}_{\mu}})\\
N_{{\rm P}} &=& N_{\nu_{\mu}}+ P_{\tau}(N_{\nu_{e}} + N_{\bar{\nu}_{\mu}})
\eea

To effectively account for the dependence on the signal normalization we use the statistic
\bea 
&\chi^{2}&  \\
&=& \sum_{i=P,D} \left(\frac{(1+\xi)N_{i,theory}- N_{i,obs}}{\sigma_{i}}\right)^{2} + \left(\frac{\xi}{\sigma_{sys}}\right)^{2} \nonumber
\eea
where the $\chi^{2}$ is minimized with respect to $\xi$, and we use $\sigma_{i}^{2} = N_{i,obs}(1+\sigma_{stat})$.   Although this is a simplified treatment of the background, we have verified that it appears to produce conservative sensitivity estimate to NSI. The dominant backgrounds we examined to CE$\nu$NS are neutrino-induced neutrons, internal $\gamma$'s and $\beta$'s, and radioactivity from the lead and concrete shielding~\cite{Akimov:2015nza}. Contributions to the systematic uncertainty include the quenching factor of Ge, the neutrino flux, and the precise knowledge of the recoil energy threshold. To account for these in a conservative manner, we take $\sigma_{stat} =0.2$ and $\sigma_{sys} = 0.1$ in the above prescription.

\subsection{Spectral Sensitivity}

Nominal sensitivity to NSI can be obtained by simply counting the total number of events  above a threshold (c.f.~\cite{Coloma:2017egw}).  Additional sensitivity is offered by a spectral analysis, which we assume should be possible. This is based on the known utility of Ge-based PPC technology at attaining excellent energy sensitivity. Based on published estimates of similar low-threshold detectors, resolutions varying from 91~${\rm eV_{ee}}$ up to 340~${\rm eV_{ee}}$ have been achieved~\cite{Soma:2014zgm}. We therefore take this pessimistic 340 ~${\rm eV_{ee}}$ value as a fiducial assumption for the resolution, which is equivalently $\approx 1.6~{\rm keV} $, where we have used the $k=2$ Lindhard paramterization the quenching factor for Ge is $(E/{\rm keV_{ee}}) = 0.2 (E_{R}/{\rm keV})^{1.12}$.

\begin{figure}[t]
\begin{center}
\includegraphics[width=.4\textwidth]{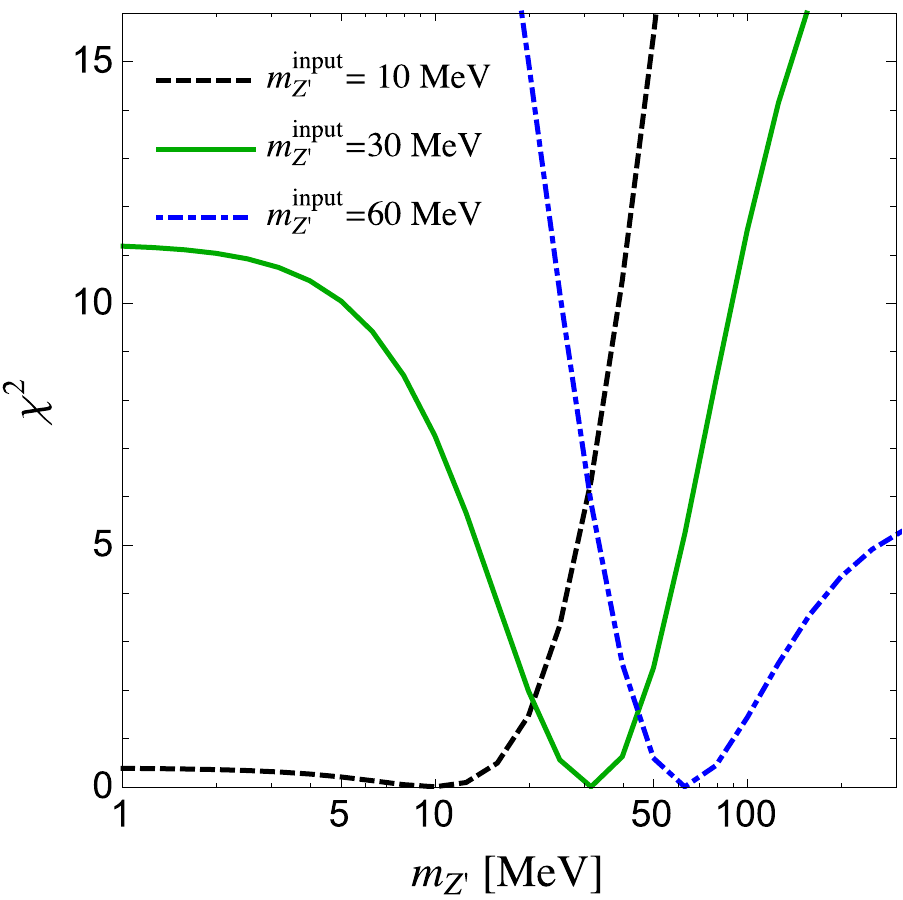}
\caption{Here we demonstrate the expected COHERENT sensitivity (2 yr of data on a 10 kg Ge target) to the finite mass of the NSI mediator from a 5 bin spectral shape analysis. To be conservative we have used a 5 keV threshold with 1.5 keV energy resolution. }
\label{fig:mass}
\end{center}
\end{figure}

\begin{figure}[t]
\begin{center}
\includegraphics[width=.43\textwidth]{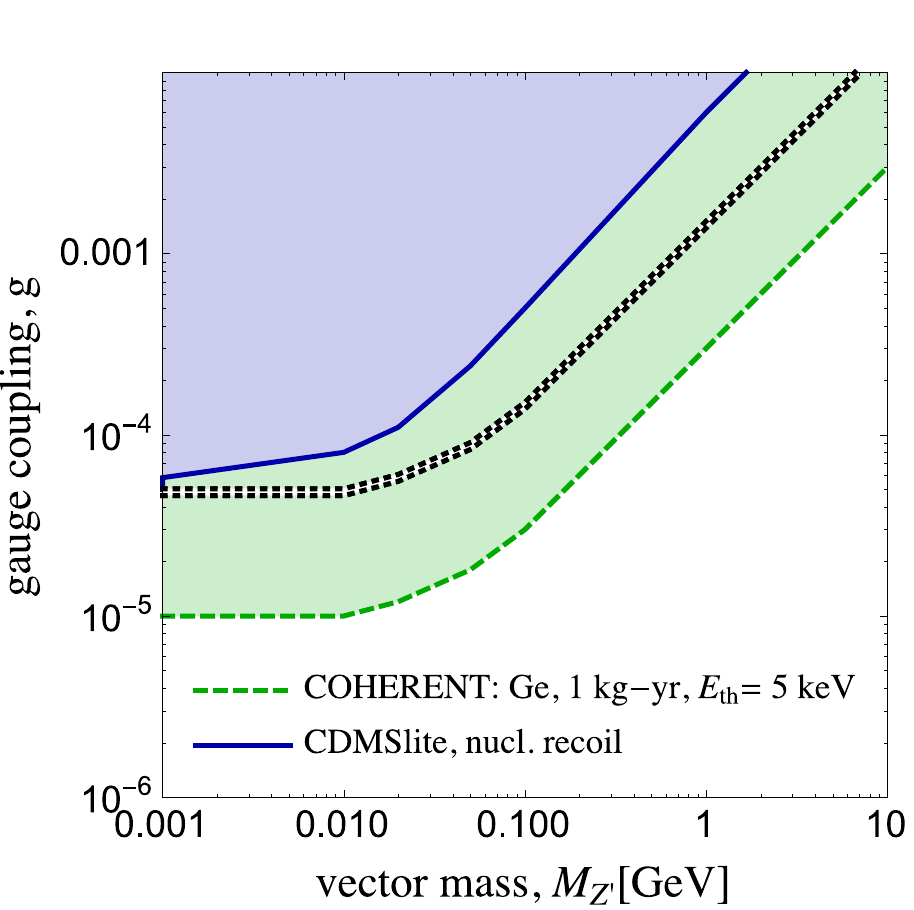}
\caption{Here in the green shaded region we display COHERENT sensitivity to new $Z'$'s interacting which couple to neutrinos and quarks universally with flavor diagonal couplings. Because of the degeneracy with the SM discussed in the text, the region between the two black dashed curves cannot be excluded.  For reference we also show the estimated limit from~\cite{Cerdeno:2016sfi} based on solar neutrino nuclear recoil interpretation of current CDMSlite data~\cite{Agnese:2015nto}.	 }
\label{fig:zprime}
\end{center}
\end{figure}

\section{Analysis Results}

We notice immediately that in addition to the expected allowed region around $ \varepsilon=0$ an interesting additional allowed region appears around $\varepsilon \simeq 0.3$. This arises from the fact that the scattering is insensitive to the sign of $Q_{\alpha\mathcal{N}}$, creating a strong degeneracy whenever the following relation is satisfied
\be 
Z \tilde{g}_{p}^{V} + (A-Z) \tilde{g}_{n}^{V} = \pm (Z g_{p}^{V} + (A-Z) g_{n}^{V}).
\ee
Thus in the vicinity of the value 
\be
\varepsilon_{ee}^{u} = - \frac{2 (Z g_{p}^{V}+(A-Z)g_{n}^{V})}{A+Z},
\ee
we indeed expect there to be limited sensitivity. This agrees quite well with the dip in sensitivity observed in the $\chi^{2}$ as it must. To some extent, the use of a detector with {\it unrenriched} Germanium (where five isotopes contribute at $>1\%$) is beneficial in moderating this degeneracy. We illustrate this in the right panel of Fig.~\ref{fig1}.  A similar degeneracy amongst NSI of different flavor couplings is also discussed in the Appendix of Ref.~\cite{Coloma:2017egw}. 

Now we will discuss our main results of an estimated sensitivity to NSI that COHERENT will have with a Ge target. These results are summarized in Fig.~\ref{fig:pheno}. Each panel assumes a different neutrino flavor coupling for NSI $\varepsilon_{\alpha \beta}$. As a result of the large fluxes of electron and muon-flavored neutrinos in the source, COHERENT will have sensitivity to all NSI coefficients with the exception of $\varepsilon_{\tau \tau}$. 

One of the most apparent features of the results is the strong sensitivity to $\varepsilon_{\mu\mu}^{f}$. This comes from a combination of two facts: (1) the $\nu_{\mu}/\bar{\nu}_{\mu}$ flux is significantly larger than the $\nu_{e}$ flux, and (2) the $\varepsilon_{\mu\mu}^{f}$ coefficient is diagonal in neutrino flavor and therefore benefits from interference with the SM contribution such that the dominant cross section modification is {\it linear} in $\varepsilon$. The sensitivity to $\varepsilon_{ee}^{f}$ is by comparison weaker as a result of the lower $\nu_{e}$ flux. Of the flavor off-diagonal NSI couplings $\varepsilon_{\alpha \beta}$ we see that $\varepsilon_{e \tau}$ is the most poorly constrained. This is easily understood since it is the only coupling which is both sensitive only the $\nu_{e}$ component of the flux and also does not benefit from interference with the SM contribution. Lastly, for the flavor off-diagonal possibilities we observe that $\varepsilon_{e \mu}$ is the most strongly constrained, as it must be since it is sensitive to all the flavors in the neutrino flux.

Next we investigate to what extent COHERENT may be sensitive to the mass scale of NSI. Deviating from the contact operator assumption (i.e. Eq~\ref{eq1}) implies that now (at least) two parameters enter into the rate: the coupling $g$ and the mediator mass $Z'$. To begin with, we will start off assuming that COHERENT find itself in the optimistic case of having discovered scattering events inconsistent with the SM. Then the question will be: what values of NSI has nature preferred?  The rough value of $\epsilon$ will be able to inferred from the total numbers of delayed and prompt events. Assuming this is the case then we can fix the combination of parameters entering into $\epsilon$ which determines the overall normalization. Thus instead of scanning over all possible values of coupling and mediator mass we instead constrain the coupling to satisfy this normalization constraint:
\be 
g^{2} = 2 \sqrt{2}~\varepsilon~G_{F}~m_{Z'}^{2}
\ee
Thus with this assumption we can carry out a subsequent more detailed analysis making use of spectral data to uncover the NSI mediator mass scale by scanning over $m_{Z'}$. This analysis is summarized in Fig.~\ref{fig:mass}. There we see that for a 10 MeV example, only a robust upper bound on the vector mass can be obtained. In contrast, for the 30 MeV and 60 MeV examples, at least a 2$\sigma$ mass measurement is possible (reaching $>3 \sigma$ in the 30 MeV case).

Lastly, in Fig.~\ref{fig:zprime} we estimate the expected COHERENT sensitivity to $Z'$ models. For reference we include an estimated bound from solar neutrino scattering in the dark matter detector CDMSlite~\cite{Agnese:2015nto} as estaimed in~\cite{Cerdeno:2016sfi}. We see that with reasonable assumptions about the exposure and detection threshold, COHERENT should be able to improve these existing limits. We also note that in this plot we compare limits derived from solar neutrino scattering~\cite{Cerdeno:2016sfi} which were obtained under the assumption that each neutrino flavor had only diagonal NSI and neutrino flavor universal couplings, i.e. $\varepsilon_{ee} = \varepsilon_{\mu\mu} = \varepsilon_{\tau\tau}$. In this case, there is no modification to the matter potential and the propagation effects from NSI vanish, leaving the only new effect as a modified CE$\nu$NS scattering rate.  In the more general case with NSI modified propagation, these limits would need to be revisited.

%
%

%

\section{Conclusions}

We have examined the near-term sensitivity of the COHERENT experiment to novel BSM contributions to neutrino-nucleus scattering. As a first step, we carefully examined the sensitivity to only one NSI parameter $\varepsilon_{ee}^{u}$.  Although the coherence of the scattering works in our favor to produce sizable cross sections, it also results in an exact degeneracy between SM scattering rates and those with NSI. This degeneracy is expectedly somewhat more pronounced for enriched germanium than for natural Ge which has a number of isotopes with sizable abundances. We anticipate that this degeneracy will be reduced further by combining data from multiple nuclei.  

Next we examined the sensitivity to each coupling NSI neutrino flavor structure separately, while assuming that the NSI new physics scale is sufficiently high that the scattering can be described entirely by dimension-6 operators.  The three flavor off-diagonal couplings $(\varepsilon_{e\mu},\varepsilon_{\mu \tau},\varepsilon_{e\tau})$ suffer reduced sensitivity from the lack of interference with the SM contribution. The coupling $\varepsilon_{\mu\mu}$ produces the strongest signals, and will therefore leads to the most stringent new constraints on NSI. This is a result of the $\nu_{e}$ flavor being the dominant contribution to the delayed signal event category. The sensitivity to $\varepsilon_{ee}$ will not be competitive however with existing constraints, which are already quite strong.  The only coupling not probed by neutrino-nucleus scattering from stopped pions is $\varepsilon_{\tau \tau}$. This is a consequence of the source-detector distance being so short compared to the oscillation length, that the source and detector and flavor ratios are equal.

Lastly, we examined a simplified renormalizable model in which NSI arises from the exchange of a light $Z'$ particle. This simplified model was studied in order to illustrate COHERENT sensitivity to We found that a coarse spectral analysis can result in useful information about the mass-scale of the $Z'$. For example if NSI arises from a $30$ MeV vector, we estimated that COHERENT can make a  mass measurement of the $Z'$ mass of $30^{+25}_{-15}$ MeV at 2$\sigma$ with 2 years of data taking on a 10 kg detector. 

Lastly, although only sensitive to NSI of the form $\varepsilon_{e\alpha}$ reactor experiments such as MINER~\cite{Agnolet:2016zir} with sensitivity to CE$\nu$NS will also be sensitive to NSI at new levels~\cite{Dutta:2015vwa}. Given the large $\bar{\nu}_{e}$ flux and likely lower energy thresholds, reactor-based CE$\nu$NS experiments may have better sensitivity to the NSI coefficients $\varepsilon_{e\alpha}$. Given the lower energy neutrino energy as well we expect the light mediator $Z'$ limits to be stronger as well. We reserve for future work a detailed examination of the complementarity between stopped-pion and reactor-based CE$\nu$NS experiments for the future of NSI.

\section*{Acknowledgements}

I am very grateful to Pilar Coloma for helpful conversations and comments on a draft of this work. I am also very grateful to Kate Scholberg and Wenqin Xu for many helpful conversations. We are also very grateful to Yuri Efremenko for comments on the first version of this work.

\vspace{2cm}

\bibliographystyle{JHEP}

\bibliography{nu}

\end{document}